\documentclass{article}

 \usepackage[preprint]{neurips_2025}

\usepackage{amsmath,amsfonts,bm}

\def\eqref#1{equation~\ref{#1}}

\def\1{\bm{1}}

\DeclareMathAlphabet{\mathsfit}{\encodingdefault}{\sfdefault}{m}{sl}
\SetMathAlphabet{\mathsfit}{bold}{\encodingdefault}{\sfdefault}{bx}{n}

\usepackage[utf8]{inputenc} 
\usepackage[T1]{fontenc}    
\usepackage{hyperref}       
\usepackage{url}            
\usepackage{booktabs}       
\usepackage{amsfonts}       
\usepackage{nicefrac}       
\usepackage{microtype}      
\usepackage{xcolor}         

\usepackage{amsmath}
\usepackage{amssymb}
\usepackage{mathtools}
\usepackage{amsthm}

\usepackage[capitalise]{cleveref}

\usepackage{microtype}
\usepackage{graphicx}
\usepackage{booktabs} 
\usepackage[textsize=tiny]{todonotes}
\usepackage{subfig}

\usepackage{wrapfig}
\usepackage{multirow}

\usepackage[utf8]{inputenc}

\title{A Benchmark for Quantum Chemistry Relaxations via Machine Learning Interatomic Potentials}

\author{
\begin{tabular}{c}
Cong Fu\textsuperscript{1}\thanks{Equal contribution}\, ,\,
Yuchao Lin\textsuperscript{1}\footnotemark[1]\, ,\,
Zachary Krueger\textsuperscript{1}\footnotemark[1]\, ,\,
Wendi Yu\textsuperscript{1},\ 
Xiaoning Qian\textsuperscript{1,2,3},\ \\
Byung-Jun Yoon\textsuperscript{2,3},\ 
Raymundo Arróyave\textsuperscript{4},\ 
Xiaofeng Qian\textsuperscript{2,4},\ 
Toshiyuki Maeda\textsuperscript{5},\\\
Maho Nakata\textsuperscript{6}\thanks{Correspondence to: Maho Nakata <maho@riken.jp> and Shuiwang Ji <sji@tamu.edu>},\hspace{0.3cm}\
Shuiwang Ji\textsuperscript{1}\footnotemark[2],\
\\
\end{tabular}\\
\textsuperscript{1}Department of Computer Science and Engineering, Texas A\&M University, USA\\
\textsuperscript{2}Department of Electrical and Computer Engineering, Texas A\&M University, USA\\
\textsuperscript{3}Computing and Data Sciences, Brookhaven National Laboratory, USA\\
\textsuperscript{4}Department of Materials Science and Engineering, Texas A\&M University, USA\\
\textsuperscript{5}Software Technology and AI Research Lab, Chiba Institute of Technology, Japan\\
\textsuperscript{6}RIKEN Cluster for Pioneering Research, RIKEN, Japan
}

\begin{document}

\maketitle

\begin{abstract}
Computational quantum chemistry plays a critical role in drug discovery, chemical synthesis, and materials science. While first-principles methods, such as density functional theory (DFT), provide high accuracy in modeling electronic structures and predicting molecular properties, they are computationally expensive. Machine learning interatomic potentials (MLIPs) have emerged as promising surrogate models that aim to achieve DFT-level accuracy while enabling efficient large-scale atomistic simulations. The development of accurate and transferable MLIPs requires large-scale, high-quality datasets with both energy and force labels. Critically, MLIPs must generalize not only to stable geometries but also to intermediate, non-equilibrium conformations encountered during atomistic simulations. In this work, we introduce PubChemQCR, a large-scale dataset of molecular relaxation trajectories curated from the raw geometry optimization outputs of the PubChemQC project. PubChemQCR is the largest publicly available dataset of DFT-based relaxation trajectories for small organic molecules, comprising approximately 3.5 million trajectories and over 300 million molecular conformations computed at various levels of theory. Each conformation is labeled with both total energy and atomic forces, making the dataset suitable for training and evaluating MLIPs. To provide baselines for future developments, we benchmark nine representative MLIP models on the dataset. Our resources are publicly available at \url{https://huggingface.co/divelab}.

\end{abstract}

\section{Introduction}
Understanding and predicting molecular behavior at the atomic scale fundamentally relies on solving the Schr\"odinger equation~\citep{griffiths2018introduction}, which governs the quantum mechanical behavior of electrons in a given atomistic system. The Schr\"odinger equation describes the electronic structure of molecules and materials by computing the wavefunction. However, due to the exponential scaling nature of electron interactions, obtaining analytical solutions is only possible for the simplest systems (\emph{e.g.}, hydrogen atom), and numerical solutions for larger systems become intractable. As a result, the development of approximate yet accurate methods for solving the electronic structure problem has become central to computational chemistry and materials science~\citep{butler2018machine,yan2024complete,zhang2023artificial}.

Density functional theory (DFT)~\citep{kohn1965self} is a widely used first-principles method that approximates the solution to the Schr\"odinger equation by modeling the electron density rather than the wavefunction directly. DFT enables the computation of quantum properties with reasonably high accuracy and has become the method of choice for quantum chemistry and atomistic simulations. However, the computational cost of DFT remains high, particularly for large systems or long timescales, limiting its applicability in high-throughput and dynamic settings. To overcome this, machine learning interatomic potentials (MLIPs)~\citep{unke2021machine} have emerged as efficient alternatives that approximate the potential energy surface (PES) learned from DFT-computed data. These models are capable of predicting total energy and atomic forces from molecular structures with near-DFT accuracy but significantly reduced computational cost. A key requirement for training accurate and transferable MLIPs is the availability of large-scale, high-quality datasets containing diverse molecular geometries annotated with energy and force labels. Importantly, for MLIPs to serve as true surrogates of DFT during molecular simulations, they must accurately model not only final, relaxed geometries, but also intermediate steps along the optimization path, which are inherently off-equilibrium. This highlights the need for datasets that contain full geometry relaxation trajectories rather than only stable structures.

In this work, we introduce PubChemQCR, a large-scale dataset of molecular relaxation trajectories curated from the raw geometry optimization data of the PubChemQC project~\citep{nakata2017pubchemqc}. To the best of our knowledge, PubChemQCR is the largest publicly available dataset of relaxation trajectories for small organic molecules, containing approximately 3.5 million trajectories and over 300 million conformations computed at various levels of theory, including 105 million conformations calculated using DFT. Each conformation is annotated with total energy and atomic force labels, which makes the dataset particularly suitable for training MLIPs and evaluating geometry optimization and molecular simulation tasks. By including both stable and intermediate geometries from actual relaxation paths, PubChemQCR offers a realistic and diverse sampling of the potential energy surface, addresses key limitations of prior datasets, such as limited element coverage, restricted conformational diversity, or the absence of force information, and enables the advancement of MLIP models for atomistic simulation.

\section{Background and Related Work}

\textbf{Machine Learning Interatomic Potentials.} Under the Born--Oppenheimer approximation~\citep{oppenheimer1927quantentheorie}, the potential energy surface (PES) of a molecular system is governed by the spatial arrangement and types of atomic nuclei. Accurately modeling this PES typically relies on quantum-mechanical approaches such as density functional theory (DFT), which are computationally intensive. Machine learning interatomic potentials (MLIPs) provide an efficient alternative by learning from DFT-generated data to predict the total energy \( E \) based on atomic coordinates \( \{\bm{x}_i\}_{i=1}^N \) and atomic numbers \( \{\bm{a}_i\}_{i=1}^N \). A standard approach expresses the total energy as a sum of atom-wise contributions, \( E = \sum_i E_i \), where each \( E_i \) is inferred from the final embedding of atom \( i \). To ensure energy conservation, atomic forces are calculated as the negative gradient of the predicted energy with respect to the atomic positions, \( \bm{f}_i = -\nabla_{\bm{x}_i} E \). By achieving near-DFT accuracy while significantly reducing computational cost, MLIPs have become widely applicable in atomistic simulations for molecular dynamics and materials modeling.


\textbf{Quantum Chemistry Datasets.} We review several related datasets in computational chemistry. QM9~\citep{ramakrishnan2014quantum} contains approximately 130,000 small molecules along with 19 quantum chemical properties. However, it includes only a single conformation per molecule, supports just 5 atom types, and does not provide atomic forces. QM7-X~\citep{hoja2021qm7} offers around 4.2 million conformations with force labels, but it covers only about 7,000 unique molecules and is limited to molecules with up to 7 heavy atoms. ANI-1x~\citep{smith2020ani} consists of over 20 million conformations spanning 57,000 unique molecules, but supports only 4 atom types. GEOM~\citep{axelrod2022geom} includes 37 million conformations across more than 450,000 molecules, yet most of its computations were performed at a lower-accuracy semi-empirical level of theory, and it lacks force labels. PubChemQC~\citep{nakata2017pubchemqc} provides ground-state structures and several quantum properties for about 3.5 million molecules, but does not release the corresponding geometry optimization trajectories. Molecule3D~\citep{xu2021molecule3d} is a large scale dataset curated from the PubChemQC for geometry prediction and ground state property prediction. PCQM4Mv2~\citep{hu2021ogb}, derived from PubChemQC and released as part of the Open Graph Benchmark (OGB)~\citep{hu2020open}, focuses on HOMO–LUMO gap prediction. MD17~\citep{bowman2022md17} and MD22~\citep{chmiela2023accurate} provide molecular dynamics (MD) trajectories for a small number of organic molecules and large molecules, respectively. OC20~\citep{chanussot2021open} and OC22~\citep{tran2023open} offer roughly 1.3 million relaxation trajectories for adsorbate–catalyst systems, but are not focused on small organic molecules. MPTrj~\citep{jain2013commentary,deng_2023_chgnet} presents optimization trajectories for materials, totaling around 1.5 million conformations. \(\nabla^2\)DFT~\citep{khrabrov2024nabla} is a comprehensive dataset containing 2 million molecules and 16 million conformations with energy, force, property, and Hamiltonian labels. However, it includes only about 60,000 trajectories and supports just 8 atom types. OMol25~\citep{levine2025open} is a large and diverse dataset comprising 83 million unique molecules and over 100 million conformations, covering small molecules, biomolecules, metal complexes, and electrolytes. There are two main kinds of geometry optimization data in OMol25. The first includes approximately 1.5 million trajectories for metal complexes, each with an average of 8 optimization steps to reach convergence. The second consists of around 5.9 million trajectories for electrolytes and multimolecular complexes, which were optimized for 2 to 5 steps to avoid overly tight relaxations.

To address the limitations of existing datasets---including restricted element coverage, limited diversity of unique molecules, and the limited availability of geometry optimization data---we introduce PubChemQCR, a new dataset curated from the raw DFT-based relaxation trajectories of the PubChemQC project~\citep{nakata2017pubchemqc}. 

\section{The PubChemQCR Dataset}

\begin{figure}[t]
\begin{center}
\centerline{\includegraphics[width=0.8\textwidth]{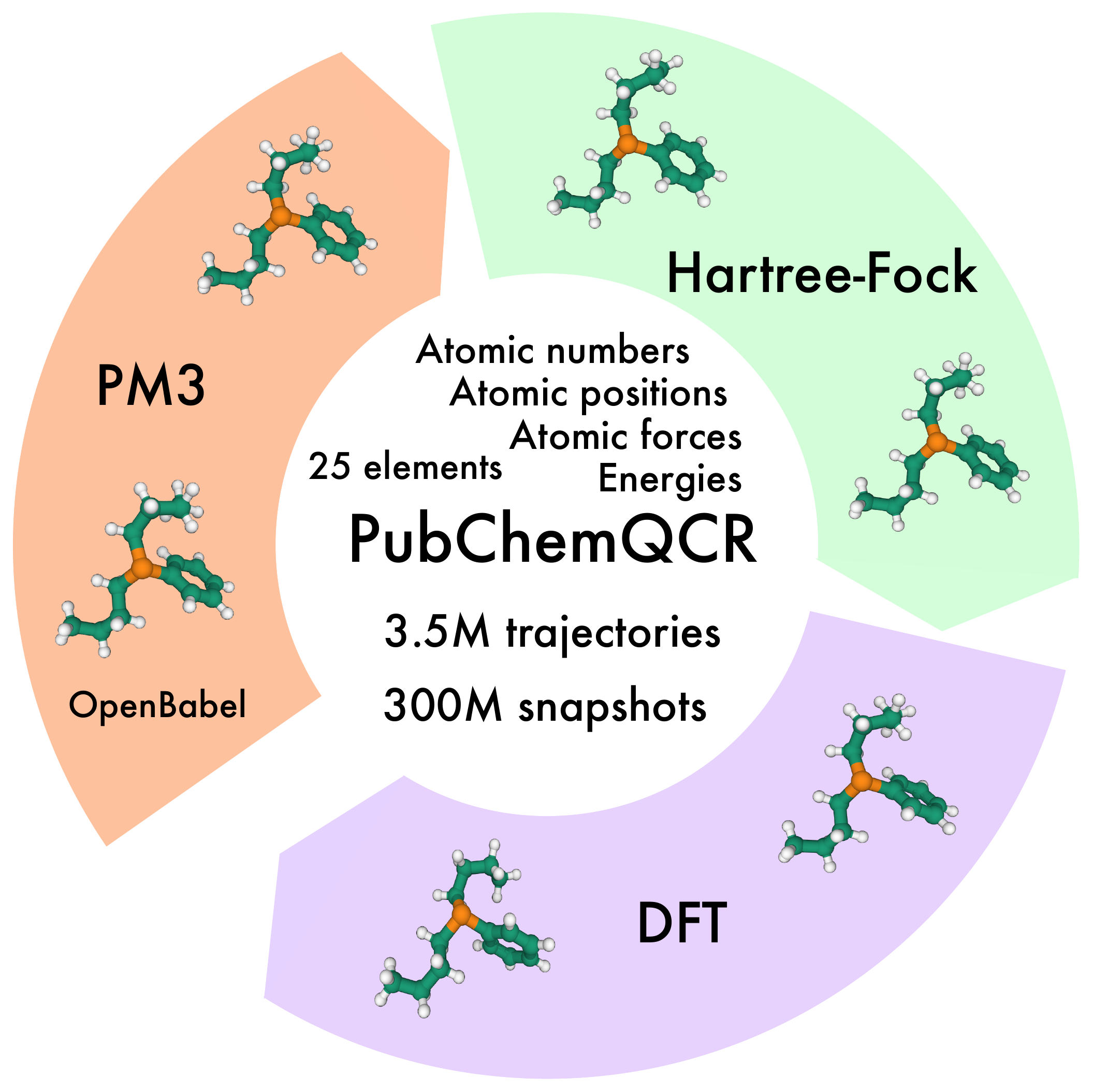}}
\caption{An overview of the PubChemQCR dataset. We curate raw geometry optimization trajectories from the PubChemQC database, where molecules are sequentially optimized using PM3, Hartree–Fock, and DFT methods. For each snapshot along the trajectory, we extract atomic numbers, atomic coordinates, atomic forces, and total energy.}
\label{fig:pipeline}
\end{center}
\vspace{-0.4in}
\end{figure}






In this section, we introduce the details about our curated PubChemQCR dataset. In~\cref{sec: data overview}, we present an overview of the dataset.~\cref{sec: data generation} describes how geometry optimizations were performed in the PubChemQC database. In~\cref{sec: data curation}, we explain the process of curating the raw geometry optimization trajectories. Finally, in~\cref{sec: data statistics}, we present key statistics of the resulting dataset.

\subsection{Overview}
\label{sec: data overview}
Obtaining stable three-dimensional conformers from initial molecular structures is a fundamental step in accurately characterizing molecular properties, as many quantum and thermodynamic properties are highly sensitive to the underlying geometry. This task is typically performed via geometry optimization, which relies on density functional theory (DFT) to compute electronic structures and their corresponding energy gradients. These gradients are then used to iteratively update atomic positions until a local minimum on the potential energy surface is reached. However, DFT-based optimization is computationally expensive, often requiring several hours to optimize a single molecule, which severely limits its scalability for high-throughput applications.

Beyond geometry optimization, accurate interatomic potentials are also critical for simulating the dynamic behavior of molecules and materials over time, as in molecular dynamics (MD). In such simulations, atomic forces must be evaluated at each time step, which, if done using DFT, becomes prohibitively expensive for large systems or long simulation times. This underscores the growing need for machine learning interatomic potentials that can serve as efficient and accurate surrogates for DFT, enabling large-scale simulations across a broad range of molecular systems.

Despite recent advances in machine learning interatomic potentials, progress has been hindered by the absence of large-scale datasets that contain high-quality DFT-level relaxation trajectories. To address this gap, we curate a new dataset, PubChemQCR, which comprises geometry optimization trajectories for approximately 3.5 million small molecules. These molecules are sourced from the PubChem Compound database, the largest publicly available repository of chemical structures represented via  IUPAC International Chemical Identifier (InChI) and Simplified Molecular Input Line Entry Specification (SMILES). The DFT-based relaxations were originally performed as part of the PubChemQC project~\citep{nakata2017pubchemqc} to obtain ground-state electronic structures. We envision that this dataset will serve as a valuable resource for training and benchmarking machine learning interatomic potentials.

\subsection{Dataset Generation}
\label{sec: data generation}
The raw trajectory data are obtained from the PubChemQC database~\citep{nakata2017pubchemqc}, a large-scale quantum chemistry resource that derives ground-state electronic structures via first-principles geometry optimization. The ground-state geometries and properties from PubChemQC were curated into the Molecule3D dataset~\citep{xu2021molecule3d}.
The geometry optimization of each molecule in PubChemQC follows a structured protocol. First, initial 3D molecular structures are generated from the InChI representation using OpenBabel~\citep{o2011open}, providing the starting point for subsequent quantum calculations.
Specifically, for the OpenBabel initialization, the 3D molecular structures were first generated using a combination of rule-based approaches and predefined ring templates. These initial geometries were then subjected to 250 steps of steepest-descent optimization using the MMFF94 interatomic potential to improve structural stability. To further explore conformational space, a weighted rotor conformer search was conducted over 200 iterations, with each candidate conformer refined through 25 steps of steepest-descent optimization. Finally, the geometries underwent an additional 250 steps of conjugate-gradient optimization to achieve better convergence and accuracy.

Following the initial geometry generation via OpenBabel, the first stage of quantum-based geometry optimization is carried out using the PM3 semi-empirical method~\citep{stewart1989optimization}, which provides a fast, approximate refinement of the molecular structure. The resulting geometries are then further optimized using the Hartree–Fock (HF) method with the STO-6G basis set to achieve a higher level of quantum accuracy. Both PM3 and HF optimizations are performed using the GAMESS~\citep{schmidt1993general} software package.

The final and most accurate stage of geometry optimization employs density functional theory (DFT) using the B3LYP functional~\citep{becke1993density} with the 6-31G* basis set. This DFT-based optimization is performed in a multi-step process to balance computational efficiency and precision. Initially, Firefly~\citep{granovsky2012firefly} or SMASH~\citep{ishimura2016scalable} is used to perform a rapid but slightly less accurate DFT geometry refinement, providing a good starting point. Subsequently, a more rigorous DFT optimization is performed using GAMESS~\citep{schmidt1993general} to ensure convergence to a local energy minimum. A final validation step is applied to confirm that each molecule is fully relaxed. This staged approach ensures high-quality geometries while optimizing computational resources.

To accelerate the computational workload, the authors of PubChemQC utilized several high-performance computing systems, including the RICC supercomputer (Intel Xeon 5570 2.93 GHz, 1024 nodes), the QUEST supercomputer (Intel Core2 L7400 1.50 GHz, 700 nodes), the HOKUSAI supercomputer (Fujitsu PRIMEHPC FX100), and the Oakleaf-FX supercomputer (Fujitsu PRIMEHPC FX10, SPARC64 IX 1.848 GHz). However, even with this extensive computational power, only a few thousand molecules can be optimized per day. Thus, it takes several years to compute the geometry optimization trajectories of 3.5 million molecules.

\subsection{Dataset Curation}
\label{sec: data curation}
The original raw trajectory data occupies approximately 7 TB of disk space and is not directly suitable for machine learning applications due to its unstructured format and redundancy. To enhance accessibility and usability, we parsed all raw log files to extract key quantum chemical properties at each optimization step, including energies, atomic forces, atomic numbers, charges, and Cartesian coordinates. Notably, DFT relaxations performed using SMASH do not provide charge information. During preprocessing, we also removed any log files that indicated failed calculations or contained duplicate logs across different optimization stages.

\begin{figure}[t]
     \begin{center}
     \subfloat[]
     {\includegraphics[width=0.49\textwidth]{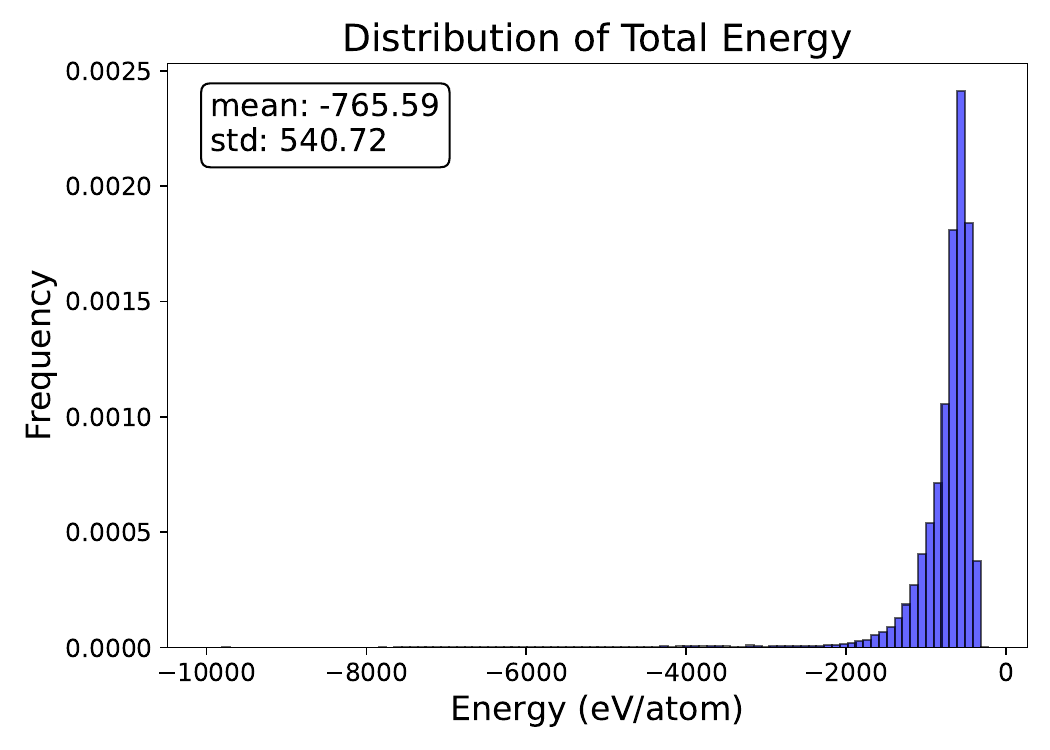}\label{fig:total_energy_distribution}}
     \subfloat[]
     {\includegraphics[width=0.49\textwidth]{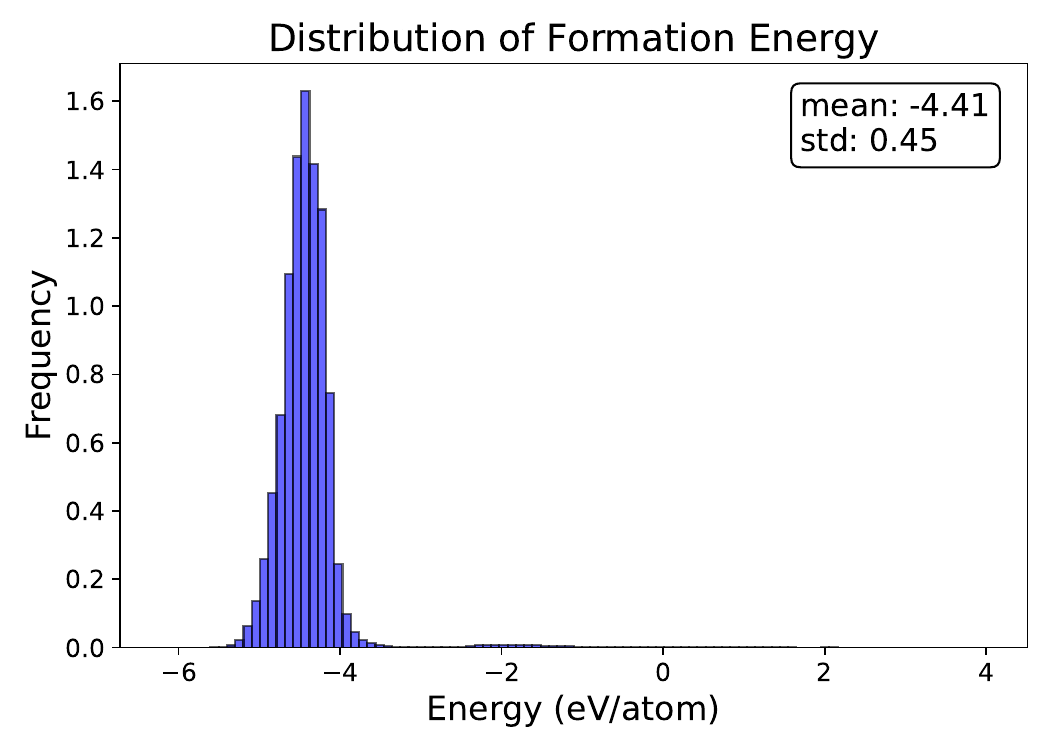}\label{fig:formation_energy_distribution}}
    \caption{Distribution of energy from the DFT optimization stage.    \protect\subref{fig:total_energy_distribution} Total energy. \protect\subref{fig:formation_energy_distribution} Formation energy obtained by subtracting the atomization energy from the total energy.
    }
    \label{fig:energy distribution}
    \end{center}
    \vspace{-0.2in} 
\end{figure}
To efficiently store and access the parsed data, we save all the parsed trajectories into Lightning Memory-Mapped Database (LMDB) files, a high-performance key–value storage format well-suited for large-scale machine learning pipelines. Each molecular trajectory is stored as a single key-value pair, where the key is the PubChem Compound ID (CID)-a unique identifier in the PubChem 
\begin{wrapfigure}{r}{0.6\textwidth}
\vspace{-0.2in}
\begin{center}
\centerline{\includegraphics[width=0.6\textwidth]{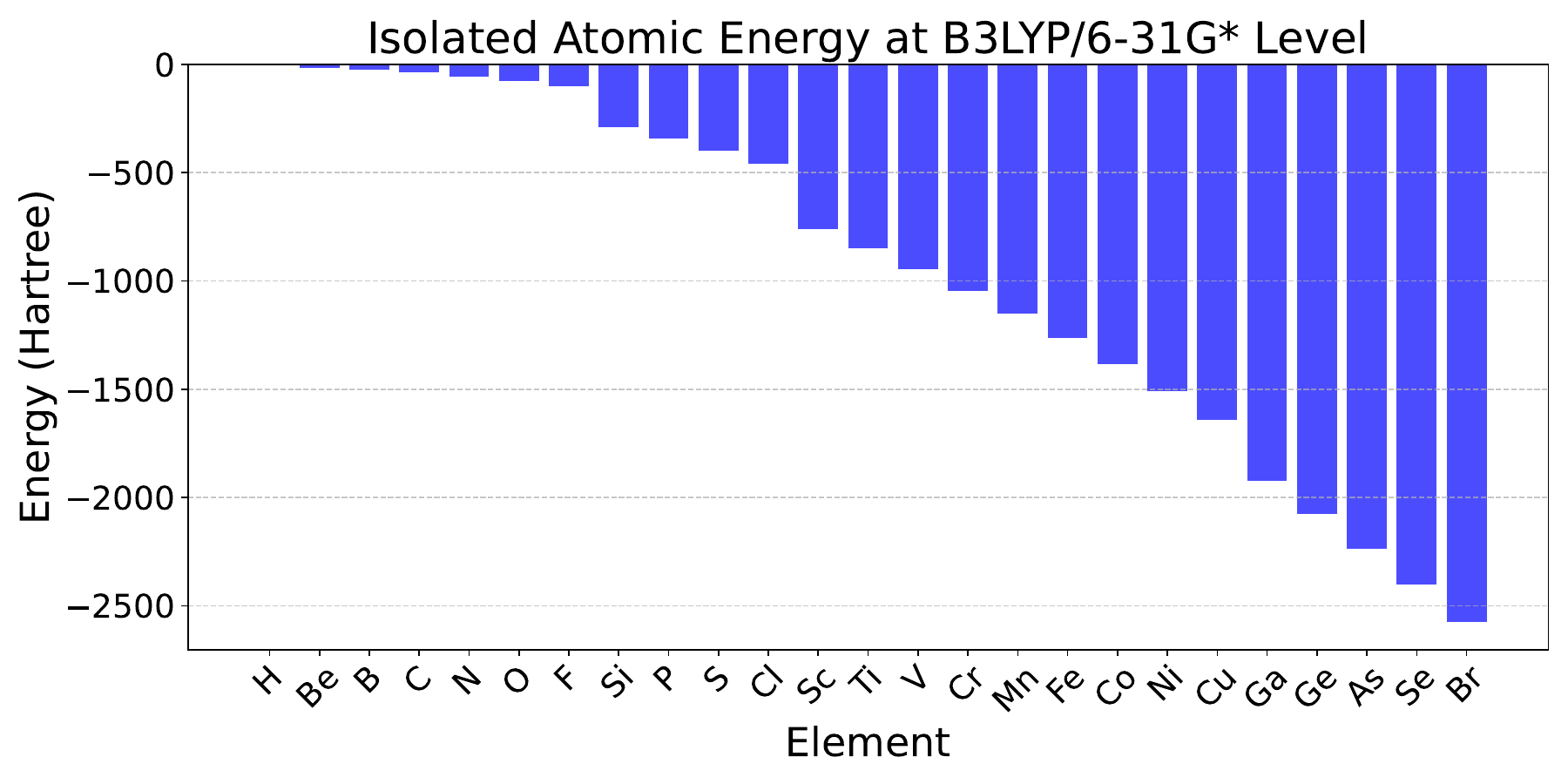}}
\vspace{-0.1in}
\caption{Isolated atomic energy calculated at DFT B3LYP/6-31G* level.}
\label{fig: atomic energy}
\end{center}
\vspace{-0.3in}
\end{wrapfigure}
database-and the value is a dictionary containing parsed data grouped by optimization stage: \texttt{pm3}, \texttt{hf}, \texttt{DFT\_1st}, and \texttt{DFT\_2nd}, where \texttt{DFT\_1st} and \texttt{DFT\_2nd} denote relaxation stage using Firefly or SMASH and GAMESS, respectively. For each method, the corresponding trajectory is represented as a list of snapshots, where each snapshot is a dictionary that stores the atomic numbers, coordinates, total energy, and force gradients, retrievable via corresponding keys. Given the challenges of training on data computed at varying levels of theory--and the fact that the DFT stage is the most critical and time-consuming--in the following, we focus exclusively on the distribution, training strategies, and benchmarks for the DFT-stage data.

Additionally, we compute isolated atomic energy for each atomic species in the dataset at the DFT B3LYP/6-31G* level of theory, as shown in~\cref{fig: atomic energy}. The isolated atomic energy refers to the total energy of a system consisting of a single, isolated atom. During training, we can normalize the molecular total energy by subtracting the atomization energies of the constituent atoms. Specifically, the target formation energy is defined as \( E_{\text{formation}} = E_{\text{total}} - \sum_{a} E_a \), where \( E_a \) denotes the isolated atomic energy of atom type \( a \) in the molecule. The energy distributions of total energy and formation energy from the DFT optimization stage are shown in~\cref{fig:energy distribution}. After subtracting the atomization energy, the resulting formation energy exhibits a mean value close to zero and a significantly reduced standard deviation. This normalization procedure removes per-atom energy offsets and reduces systematic bias, leading to a more compact and centered energy distribution. As a result, the learning task becomes easier and the model tends to converge more efficiently during training.

In summary, by organizing snapshots according to their respective optimization stages, the dataset enables flexible selection of trajectory segments for targeted training scenarios. The final curated dataset has been compressed to approximately 400 GB, a significant reduction from the original size, while preserving all essential information. To further facilitate adoption, we provide a customized PyTorch Geometric (PyG)~\citep{fey2019fast} dataloader that is fully compatible with geometric deep learning models. This combination of compact storage, structured access, and ready-to-use tooling significantly lowers the barrier for researchers to experiment with and benchmark machine learning interatomic potentials on realistic quantum trajectories.

\subsection{Dataset Statistics}
\label{sec: data statistics}
\begin{wrapfigure}{r}{0.45\textwidth}
\vspace{-0.2in}
\begin{center}
\centerline{\includegraphics[width=0.45\textwidth]{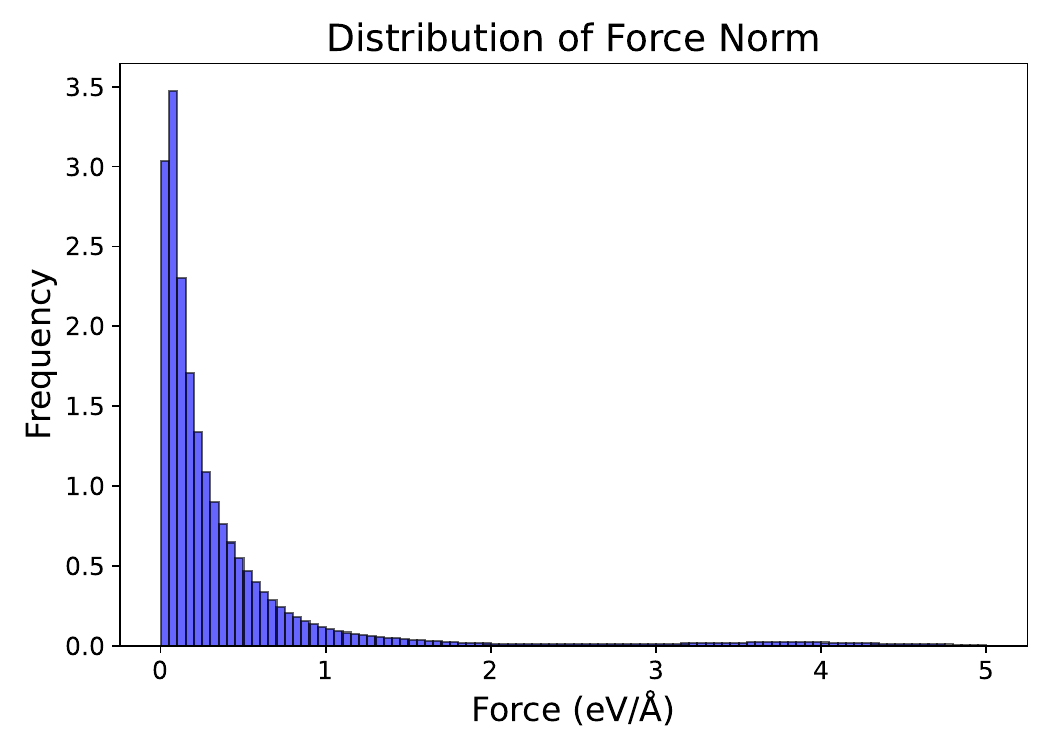}}
\caption{Distribution of forces from the DFT optimization stage.}
\label{fig: force distribution}
\end{center}
\vspace{-0.2in}
\end{wrapfigure}
The full PubChemQCR dataset comprises 3,471,000 molecular relaxation trajectories and a total of 298,751,667 molecular snapshots spanning multiple levels of quantum chemical theory. Specifically, as shown in~\cref{tab: data statistics}, it includes 163,015,359 snapshots from PM3, 19,274,130 snapshots from Hartree–Fock, 105,494,671 snapshots from the first substage of DFT optimization, and 10,967,507 snapshots from the second substage of DFT. On average, each molecule consists of 29 atoms, including 14 heavy atoms, and each trajectory contains approximately 47 PM3 snapshots, 6 Hartree–Fock snapshots, 31 DFT (first substage) snapshots, and 3 DFT (second substage) snapshots. The semi-empirical PM3 and ab initio Hartree–Fock stages are considerably more efficient, typically requiring only a few minutes per molecule, whereas DFT optimizations can take several hours. The force distribution of the DFT optimization stage is shown in~\cref{fig: force distribution}. The dataset spans 25 chemical elements, including H, Be, B, C, N, O, F, Si, P, S, Cl, Sc, Ti, V, Cr, Mn, Fe, Co, Ni, Cu, Ga, Ge, As, Se, and Br, providing broad chemical diversity for training and evaluating machine learning models on realistic molecular systems.

To facilitate efficient model development and rapid benchmarking, we further curated a smaller subset of the full dataset, referred to as PubChemQCR-S. This subset comprises 40,979 molecular relaxation trajectories and 1,504,431 DFT-calculated snapshots, specifically selected from the first substage of the DFT optimization process. PubChemQCR-S is ideal for use in ablation studies, hyperparameter tuning, and preliminary evaluations of machine learning interatomic potentials.

\begin{table}[t]
  \centering  
  \caption{Total number of snapshots and average number of snapshots per molecule at each stage of geometry optimization, after curating the raw trajectory data from PubChemQC.}
  \vspace{0.1in}
  \begin{tabular}{lcc}
    \toprule
    Stages  &  \#Snapshots & \#Avg Snapshots per Molecule \\
    \midrule
    Total & $298,751,667$ & $87$ \\
    \midrule
    PM3     & $163,015,359$  & $47$   \\
    Hartree Fock     &  $19,274,130$    & $6$    \\
    DFT (Firefly/SMASH)   & $105,494,671$   & $31$  \\
    DFT (GAMESS)    & $10,967,507$   & $3$  \\
    \bottomrule
  \end{tabular}
  \label{tab: data statistics}
\end{table}


\section{Benchmarking}
In~\cref{sec: energy_force_prediction}, we benchmark the energy and force prediction performance of representative MLIP models. In~\cref{sec:geometry optimization}, we evaluate the geometry optimization capabilities of MLIP models trained on the PubChemQCR-S dataset.

\subsection{Energy and Force Prediction}
\label{sec: energy_force_prediction}
\textbf{Task.} Machine learning interatomic potentials aim to predict the total energy and atomic forces from a given three-dimensional molecular structure, which consists of atomic numbers and corresponding 3D coordinates. Atomic forces can be obtained either by computing the negative gradient of the predicted energy with respect to atomic positions or by employing a separate prediction head that directly predicts the force vectors.

\textbf{Dataset Splits.} For the PubChemQCR-S subset, we partition the data into training, validation, and test sets using a 60\%/20\%/20\% split. For the full PubChemQCR dataset, we retain the same test set as used in the PubChemQCR-S subset, and divide the remaining data into training and validation sets with an 80\%/20\% split. To ensure data integrity and avoid information leakage, each geometry optimization trajectory is assigned exclusively to a single split in both the PubChemQCR-S and full PubChemQCR datasets. Note that we only use data from the DFT optimization stage to train the model. 

\textbf{Baseline Methods.} We benchmark several representative machine learning interatomic potential baselines on the PubChemQCR-S subset:

\begin{itemize}
\item \textbf{SchNet}~\citep{schutt2018schnet}: An invariant model that utilizes continuous-filter convolutional networks to capture local atomic environments through filters generated by learned networks. 
\item \textbf{PaiNN}~\citep{schutt2021equivariant}: An equivariant model that advances SchNet by integrating equivariant feature representations. PaiNN effectively captures directional information in molecular systems, enabling accurate predictions of both scalar and tensorial properties.
\item \textbf{FAENet}~\citep{duval2023faenet}: An equivariant model that uses frame averaging techniques to ensure equivariance symmetry of molecule structures through data transformations while avoiding symmetry-preserving architectural constraints.
\item \textbf{NequIP}~\citep{batzner20223}: A model that uses equivariant convolutions for interactions of geometric tensors. Specifically, the model encodes the atomic environments faithfully through modeling the feature interactions via the Clebsch-Gordan tensor product.
\item \textbf{SevenNet}~\citep{park2024scalable}: An $E(3)$-equivariant model that extends the NequIP architecture with a scalable parallelization algorithm tailored for spatial decomposition in large-scale molecular dynamics (MD) simulations. 
\item \textbf{Equiformer}~\citep{liao2022equiformer}: This model extends Transformers to 3D molecular graphs by replacing the standard operations with their equivariant counterparts based on tensor products and by designing graph attention mechanisms that preserve geometric equivariance.
\item \textbf{Allegro}~\citep{musaelian2023learning}: A scalable and computationally efficient local equivariant model that maintains equivariance and achieves high accuracy by constructing many-body interactions through a hierarchy of tensor products based on learned equivariant features.
\item \textbf{MACE}~\citep{batatia2022mace}: This model addresses the limitation of traditional message-passing neural networks, which rely solely on two-body interactions, by integrating equivariant message passing with higher-order body interactions. 
\item \textbf{PACE}~\citep{xu2024equivariant}: An equivariant graph network that utilizes edge
booster and the Atomic Cluster Expansion (ACE) techniques to approximate the equivariant polynomial functions for atomic energy and force prediction.
\end{itemize}

Due to the scale of the full PubChemQCR dataset and the associated computational cost, we restrict our benchmarks to small and computationally efficient models. Specifically, we evaluate the performance of FAENet~\citep{duval2023faenet} and PaiNN~\citep{schutt2021equivariant} on the full dataset.

\textbf{Training Setup.} On the PubChemQCR-S subset, Equiformer uses a separate prediction head to directly predict atomic forces, whereas other methods compute forces as the gradient of the predicted energy. In the complete PubChemQCR dataset, PaiNN also employs a separate force prediction head. Note that to eliminate the influence of molecular size, we predict the energy per atom rather than the total energy. Additionally, we normalize the energy by subtracting the mean energy during training. To remove the effect of translation, we also center the coordinates by shifting them to have a zero centroid.
\begin{table}[t]
  \centering  
  \caption{Model configurations—including the number of layers, hidden dimensions (or maximum irreducible representation channels), and batch sizes—are provided for all baseline models trained on the PubChemQCR-S dataset. These models include SchNet~\citep{schutt2018schnet}, PaiNN~\citep{schutt2021equivariant}, MACE~\citep{batatia2022mace}, Equiformer~\citep{liao2022equiformer}, PACE~\citep{xu2024equivariant}, FAENet~\citep{duval2023faenet}, NequIP~\citep{batzner20223}, 
  SevenNet~\citep{park2024scalable}, and Allegro~\citep{musaelian2023learning}.}
  \vspace{0.1in}
  \begin{tabular}{lccc}
    \toprule
    Models      & Layers & Hidden Dimension & Batch Size \\
    \midrule
    SchNet              & 4               & 128                  & 128                 \\
    PaiNN               & 4               & 128                  & 32                 \\
    FAENet              & 4               & 128                  & 64                 \\
    NequIP              & 5               & 64                  & 16                  \\
    SevenNet            & 5               & 128                  & 16                  \\
    MACE                & 2              & 128                  & 8                  \\
    PACE                & 2               & 128                  & 8                  \\
    Allegro             & 2               & 128                  & 8                 \\
    Equiformer                & 4               & 128                  & 32                  \\

    \bottomrule
  \end{tabular}
  \label{tab:baseline-configs}
\end{table}

\cref{tab:baseline-configs} summarizes the model configurations used for all baseline methods. For FAENet, we adopt the “simple” message-passing variant, while for MACE, we include the residual interaction block to enhance expressiveness. Initial attempts to train Equiformer using its original OC20 settings (6 layers with hidden irreps of either 256$\times$0e + 256$\times$1e or 256$\times$0e + 128$\times$1e) failed to converge; thus, we employ a reduced configuration consisting of 4 layers with irreps 128$\times$0e + 64$\times$1e. For all tensor-product-based models—including NequIP~\citep{batzner20223}, MACE~\citep{batatia2022mace}, PACE~\citep{xu2024equivariant}, Allegro~\citep{musaelian2023learning}, SevenNet~\citep{park2024scalable}, and Equiformer~\citep{liao2022equiformer}—only even-parity irreducible representations are used and $L_{max} = 2$ except for Equiformer.

All experiments on the PubChemQCR benchmarks employ a cutoff radius of $4.5\,\text{\AA}$, the Adam optimizer~\citep{kingma2014adam} with an initial learning rate of $5\times10^{-4}$, and a \textsc{ReduceLROnPlateau} learning rate scheduler with a patience of 2 epochs. Models are trained for up to 100 epochs on the PubChemQCR-S subset and up to 15 epochs on the full PubChemQCR dataset, using NVIDIA A100-80GB GPUs. 

\textbf{Evaluation Metrics.} We measure the accuracy of the machine learning interatomic potential prediction by calculating the mean absolute error (MAE) of energies and the root mean square error (RMSE) of forces, shown as below:
\begin{align}
    \mathcal{L}_{\textsubscript{MAE}} &= \frac{1}{N}\sum_{i=0}^N |\hat{e_i} - e_i|, \\
    \mathcal{L}_{\textsubscript{RMSE}} &= \sqrt{\frac{1}{M} \sum_{i=0}^M ||\bm{f}_i - \hat{\bm{f}_i}||^2},
\end{align}
where $\hat{e_i}$ represents the predicted energy and $e_i$ is the ground truth energy. Similarly, $\hat{\bm{f}_i} \in \mathbb{R}^3$ denotes the predicted forces for each atom, while $\bm{f}_i$ represents the ground truth forces.

\textbf{Results and Discussions.} The results of benchmarking on PubChemQCR-S are shown in~\cref{tb:small}. Equiformer achieves the best overall performance, with the lowest force MAE on both validation and test sets. It also yields the best energy MAE on the validation set and a competitive energy MAE on the test set. PaiNN ranks second overall, showing strong performance in both energy and force prediction. SchNet, although older and less expressive than some equivariant models, still performs reasonably well, particularly in energy prediction. Models like MACE, PACE, and NeuqIP show moderate performance, outperforming the worst-performing models but not reaching the level of Equiformer or PaiNN. FAENet and Allegro perform poorly on this dataset compared to others, possibly due to their architectural assumptions or overfitting. SevenNet achieves comparable force prediction performance to PaiNN but exhibits much worse energy prediction accuracy. The benchmarking results on the full PubChemQCR dataset are shown in~\cref{tb:full}. PaiNN and FAENet achieve improved performance when trained on the full dataset compared to the smaller subset, highlighting the importance of large-scale data for training accurate machine learning interatomic potentials.

\begin{table}[t]
    \centering
    \caption{Comparison of model performance on energy and force prediction tasks using the PubChemQCR-S dataset. The best results are highlighted in bold.}
        \vspace{0.05in}
    \label{tb:small}
    \resizebox{0.8\textwidth}{!}{  
\begin{tabular}{lccccc}
\toprule
                      & \multicolumn{2}{c}{Validation}                & \multicolumn{2}{c}{Test} \\
 \multirow{2}{*}{Model}      & {Energy MAE}          & {Force RMSE}           & {Energy MAE}          & {Force RMSE} \\ 
                        &                          (meV/atom) $\downarrow$    & (meV/\AA) $\downarrow$& (meV/atom) $\downarrow$    & (meV/\AA) $\downarrow$\\
 \midrule
      
         SchNet                  & $5.30$  & $56.55$ &  $5.55$ & $56.22$\\
         PaiNN                  & $5.13$ & $46.34$ & $\textbf{5.33}$ & $46.92$ \\
         NequIP    & $7.37$ & $54.78$ & $8.27$ & $55.59$ \\
         SevenNet   & $8.77$ & $47.63$ & $10.21$ & $48.05$ \\
         Allegro   & $10.86$ & $60.71$ & $10.80$ & $61.44$ \\
         FAENet   & $7.28$ &  $60.24$ & $8.70$ & $60.51$ \\
         MACE                   & $7.54$ & $51.46$ & $7.47$ & $45.70$ \\
         PACE                  & $6.24$ & $50.54$ & $6.53$ & $51.43$ \\
         Equiformer                   & $\mathbf{4.69}$ & $\mathbf{34.67}$ & $5.38$ & $\mathbf{35.11}$ \\

    \bottomrule
\end{tabular}
}
\end{table}

\begin{table}[t]
    \centering
    \caption{Comparison of model performance on energy and force prediction tasks using the full PubChemQCR dataset. The best results are highlighted in bold.}
        \vspace{0.05in}
    \label{tb:full}
    \resizebox{0.8\textwidth}{!}{  
\begin{tabular}{lccccc}
\toprule
                      & \multicolumn{2}{c}{Validation}                & \multicolumn{2}{c}{Test} \\
 \multirow{2}{*}{Model}      & {Energy MAE}          & {Force RMSE}           & {Energy MAE}          & {Force RMSE} \\ 
                        &                          (meV/atom) $\downarrow$    & (meV/\AA) $\downarrow$& (meV/atom) $\downarrow$    & (meV/\AA) $\downarrow$\\

\midrule
  FAENet & $4.86$ & $59.60$ & $6.16$ & $51.00$\\
  PaiNN  & $\mathbf{2.47}$ & $\mathbf{36.39}$ & $\mathbf{1.91}$ & $\mathbf{23.86}$ \\
    \bottomrule
\end{tabular}
}
\end{table}

\subsection{Geometry Optimization}
\label{sec:geometry optimization}
\textbf{Task.} The goal of geometry optimization is to iteratively update atomic positions in order to minimize the system's potential energy to obtain stable 3D geometries.

\textbf{Dataset.} To ensure high structural diversity, we sample 1,000 molecules from the PubChemQCR-S test set using the MaxMin diversity strategy. We begin by computing Morgan fingerprints for all molecules in the test set. An initial molecule is randomly chosen, and subsequent selections are made by iteratively identifying the molecule with the greatest Tanimoto distance from the already selected set.

\textbf{Simulation Protocol. } Since the baseline models are trained on data from the DFT relaxation stage, we perform geometry optimization starting from the first snapshot of the DFT stage (i.e., the structure obtained after the PM3 and Hartree–Fock stages). We use the ASE~\citep{larsen2017atomic} package to perform geometry optimization. Specifically, we implement a custom calculator for each trained MLIP model to predict atomic forces for a given molecular structure. The Broyden–Fletcher–Goldfarb–Shanno (BFGS) algorithm~\citep{fletcher2000practical} is employed to iteratively update atomic coordinates. The maximum step size is set to 0.2\AA. The optimization process terminates when either the maximum atomic force falls below \( 0.05 \, \text{eV/Å} \), or the number of optimization steps exceeds 500.

\textbf{Evaluation Metrics. } Our evaluation metrics are adapted in part from those introduced in~\citep{tsypin2023gradual}, and include the following:  
(1) \textbf{Average Energy Minimization Percentage}, $\overline{\mathrm{pct}}_T$, which captures the extent to which MLIP-based optimization reduces the system energy relative to DFT-optimized geometries;  
(2) \textbf{Chemical Accuracy Success Rate}, $\mathrm{pct}_{\text{success}}$, defined as the fraction of molecules whose final energy lies within the chemical accuracy threshold (typically 1 kcal/mol);  
(3) \textbf{Divergence Rate}, $\mathrm{pct}_{\text{div}}$, indicating the proportion of molecules for which either the DFT single-point energy calculation fails or the final DFT energy exceeds the starting value;  
(4) \textbf{Force Convergence Rate}, $\mathrm{pct}_{\text{FwT}}$, which reports the percentage of cases where the maximum atomic force is below 0.05 eV/\AA{} after optimization. Additional details are provided in~\cref{app: geometry optimization metrics}.


\textbf{Results and Discussions.} The results of geometry optimization is shown in~\cref{tab: geometry optimization}. Since the MLIP models are trained on near-optimal relaxation data, the atomic forces in these configurations are relatively small and subtle, making it essential for the models to learn highly accurate force predictions for effective geometry optimization. Among all models, Equiformer achieves the best performance, attaining an average energy minimization percentage of 70.15\%, a chemical accuracy success rate of 23.81\%, and a relatively low divergence rate of 5.26\%. Furthermore, it exhibits a notable force convergence rate, reaching 19.85\% of cases below the 0.05 eV/\AA{} threshold. This indicates that Equiformer is more effective at capturing fine-grained geometric gradients and optimizing molecular structures when using near-equilibrium training data.
In contrast, most other models have substantially worse performance than Equiformer. For instance, SevenNet, despite its relatively strong performance in force prediction tasks, exhibits poor optimization outcomes with an 1.45\% force convergence rate and a divergence rate of 18.51\%, indicating frequent failures to maintain stability. 
Moreover, models such as SchNet, Allegro, and FAENet achieve moderate success in energy minimization, with $\overline{\mathrm{pct}}_T$ ranging between 40\% and 50\%, but all exhibit near-zero force convergence, further reinforcing the difficulty of driving relaxation from near-optimal initial states.


\begin{table}[t]
  \caption{Geometry optimization performance of the MLIP models trained on PubChemQCR-S.}
  \vspace{0.05in}
  \label{tab: geometry optimization}
  \centering
  \resizebox{0.8\textwidth}{!}{
  \begin{tabular}{lcccc}
    \toprule
    Model     &  $\overline{\mathrm{pct}}_T$(\%)$^{\uparrow}$ & $\mathrm{pct}_{\text{success}}$ (\%)$^{\uparrow}$ & $\mathrm{pct}_{\text{div}}$ (\%)$^{\downarrow}$& $\mathrm{pct}_{\text{FwT}}$ (\%)$^{\uparrow}$\\
    \midrule
    PaiNN & $54.83$ & $7.62$ & $9.85$ &  $2.03$ \\
    SchNet & $50.97$ & $4.43$ & $26.42$ & $0.10$    \\
    Equiformer & $\mathbf{70.15}$ & $\mathbf{23.81}$ & $\mathbf{5.26}$ & $\mathbf{19.85}$ \\
    NequIP & $47.02$ & $4.68$ & $22.00$ & $1.31$    \\
    SevenNet & $45.93$ & $8.86$ & $18.51$ & $1.45$ \\
    Allegro & $42.09$ & $2.10$ & $15.87$ & $0.0$    \\
    FAENet & $45.94$ & $1.79$ & $21.3$ & $0.0$    \\
    MACE & $50.15$ & $5.70$ & $5.54$ & $0.61$    \\
    PACE & $47.98$ & $5.75$ & $19.90$ & $0.69$  \\
    \bottomrule
  \end{tabular}
  }
\end{table}

\section{Potential Applications}
\textbf{MLIP Training for Efficient Atomistic Simulations.} Simulating atomistic systems plays a critical role in understanding the dynamic behavior of molecular and material systems~\citep{best2012atomistic,dove2008introduction,bernstein2009hybrid}, with broad applications in drug discovery, biological research, and materials science. Traditionally, such simulations rely on \emph{ab initio} molecular dynamics using density functional theory (DFT) or on empirical interatomic potentials. However, DFT-based simulations are computationally intensive, and empirical models often lack accuracy and generalizability. Machine learning interatomic potentials (MLIPs)~\citep{unke2021machine,wang2024machine} offer an attractive alternative, serving as surrogate models that approximate DFT-level accuracy while significantly reducing computational cost. MLIPs trained on our dataset can potentially be applied to perform molecular dynamics simulations with improved efficiency~\citep{lin2024equivariance,wang2022comenet,liu2021spherical}.

Furthermore, obtaining stable and reliable 3D molecular geometries is essential for accurate prediction of quantum properties in small molecules~\citep{chen2024geomformer}. However, this typically requires costly geometry optimization procedures involving repeated DFT calculations~\citep{nakata2017pubchemqc}. Since our dataset contains full geometry optimization trajectories, MLIPs trained on it can serve as replacements for DFT in geometry optimization workflows. This enables rapid and cost-effective generation of low-energy molecular conformations, thereby facilitating downstream tasks in quantum chemistry and molecular design.

\textbf{Pre-Training for Downstream Property Prediction.} When training machine learning interatomic potentials (MLIPs), models learn to approximate energies and atomic forces, effectively capturing fundamental interatomic interactions. Through this process, the model acquires rich, physically meaningful representations of molecular systems. In molecular representation learning~\citep{wang2019smiles,liu2021fast,liu2021pre,stark20223d}, the goal is to obtain latent representations that encode structural and chemical information, which can be leveraged for a variety of downstream property prediction tasks.

While self-supervised learning~\citep{hu2019strategies, rong2020self,zaidi2022pre, feng2023fractional, ni2023sliced, liao2024generalizing} is commonly employed on large unlabeled molecular datasets to learn transferable representations, our curated dataset provides explicit energy and force labels, enabling supervised pre-training, which allows the model to learn physically grounded representations. These pre-trained models can then be fine-tuned on downstream tasks such as quantum property prediction, potentially improving both accuracy and generalization.

\textbf{Training 3D Molecular Generative Models.} Our dataset's comprehensive collection of DFT-optimized geometry trajectories enables the development of advanced 3D generative models capable of directly synthesizing molecular structures in three-dimensional space~\citep{xu2022geodiff,xu2023geometric,hoogeboom2022equivariantdiffusionmoleculegeneration,fu2024latent,fu2024fragment}. These models offer significant advantages over traditional 2D graph-based approaches~\citep{shi2020graphafflowbasedautoregressivemodel,luo2021graphdfdiscreteflowmodel} by eliminating the need for separate conformer generation processes and instead learning to sample energetically stable, physically realistic geometries in an end-to-end fashion. This direct 3D generation capability proves essential for applications in pharmaceutical design, materials science, and catalysis research, where precise three-dimensional conformations dictate molecular functions.

The dataset's rich sampling of conformational landscapes around energy minima, combined with accurate energy and force annotations, provides robust training signals for physics-informed generative architectures. Through training on DFT-validated structures, such models learn to generate chemically sound, low-energy molecular conformations ready for immediate application with minimal optimization steps.

\section{Limitations}

In PubChemQC database, geometry optimization is performed in a sequential manner—initially with PM3, followed by Hartree–Fock, and finally with density functional theory (DFT). As a result, the energy and force labels obtained at different stages of the same trajectory exhibit varying levels of accuracy and are neither directly comparable nor mutually consistent. Consequently, when training machine learning interatomic potentials (MLIPs), it is advisable to utilize only the DFT-optimized segments, as they provide the highest fidelity labels. However, this restriction limits the dataset to the near-equilibrium region of the potential energy surface, where atomic forces tend to be small. This poses a significant challenge for training MLIPs, which require diverse force magnitudes and broader sampling of molecular conformational space for generalization.

Additionally, the dataset includes only 25 chemical elements, constrained by the compatibility of the chosen DFT functional and basis set. To improve the dataset's coverage of chemical space, it is essential to incorporate a wider range of elements and more diverse atomic interactions. Furthermore, to ensure label consistency throughout the optimization trajectory, it would be preferable to perform the entire geometry optimization using a single, uniform level of DFT functional and basis set.

\section{Summary} 
In this work, we introduce PubChemQCR, the largest publicly available dataset of DFT-based relaxation trajectories for small organic molecules. Comprising approximately 3.5 million trajectories and over 300 million conformations---with 105 million computed using DFT---PubChemQCR provides high-quality energy and force labels essential for training and evaluating machine learning interatomic potentials. By addressing limitations in existing datasets related to element coverage, conformational diversity, and trajectory availability, PubChemQCR enables the development of more accurate and transferable models for atomistic simulation and molecular property prediction.

\section*{Acknowledgments}

XNQ, BY, and SJ acknowledge support from ARPA-H under grant 1AY1AX000053. SJ acknowledges partial support from the National Institutes of Health under grant U01AG070112 and National Science Foundation under grant IIS-2243850. XNQ acknowledges partial support from National Science Foundation under Grants SHF-2215573 and IIS-2212419. RA acknowledges support from National Science Foundation under grant 2119103. XFQ acknowledges support from the National Science Foundation under grant DMR-1753054 and partial support by the donors of ACS Petroleum Research Fund under grant \#65502-ND10.

\bibliography{cong,dive,FM}
\bibliographystyle{unsrtnat}

\newpage
\appendix

\section{Appendix}
\subsection{Metrics for Geometry Optimization}
\label{app: geometry optimization metrics}
\textbf{Average Energy Minimization Percentage.} This metric evaluates the extent to which the energy is reduced by the MLIP-optimized structure when compared to the DFT-optimized geometry:
\begin{align}
    \overline{\mathrm{pct}}_T = \frac{1}{|\mathcal{D}_{\text{opt}}|} \sum_{c \in \mathcal{D}_{\text{opt}}} \mathrm{pct}(c_T),
\end{align}
where the individual percentage improvement \( \mathrm{pct}(c_T) \) is calculated as:
\begin{align}
    \mathrm{pct}(c_T) = 100\% \cdot \frac{E_{c_0}^{\mathrm{DFT}} - E_{c_T}^{\mathrm{DFT}}}{E_{c_0}^{\mathrm{DFT}} - E_{c_{\mathrm{gt}}}^{\mathrm{DFT}}}.
\end{align}
Here, \( c_0 \), \( c_T \), and \( c_{\mathrm{gt}} \) correspond to the starting conformer, the structure optimized by MLIP, and the ground truth DFT-optimized conformer, respectively. The notation \( E^{\mathrm{DFT}}_{(\cdot)} \) refers to the DFT-calculated energy at a given geometry. When calculating this metric, we only consider optimized conformers with $E_{c_T}^{\mathrm{DFT}}$ lower than $E_{c_0}^{\mathrm{DFT}}$. If $E_{c_T}^{\mathrm{DFT}}$ is lower than $E_{c_{\mathrm{gt}}}^{\mathrm{DFT}}$, we only consider it when the force calculated by DFT is within the threshold, indicating that the model confidently finds a conformer with lower energy than DFT-based relaxation.

\textbf{Chemical Accuracy Success Rate.} This metric quantifies the proportion of MLIP-relaxed geometries that achieve a residual energy below the standard threshold of 1 kcal/mol:
\begin{align}
    \mathrm{pct}_{\text{success}} = \frac{1}{|\mathcal{D}_{\text{opt}}|} \sum_{c \in \mathcal{D}_{\text{opt}}} I\left[ E^{\mathrm{res}}(c_T) < 1 \right],
\end{align}
where the residual energy is defined as:
\begin{align}
    E^{\mathrm{res}}(c_T) = E_{c_T}^{\mathrm{DFT}} - E_{c_{\mathrm{gt}}}^{\mathrm{DFT}}.
\end{align}

\textbf{Divergence Rate.} This metric, denoted as $\mathrm{pct}_{\text{div}}$, captures the fraction of cases where the DFT single-point energy evaluation fails or the energy of the optimized structure exceeds that of the initial geometry.

\textbf{Force Convergence Rate.} This metric determines how often the MLIP-relaxed conformers meet the force convergence criterion, specifically having the maximum atomic force fall below 0.05 eV/\AA:
\begin{align}
    \mathrm{pct}_{\text{FwT}} = \frac{1}{|\mathcal{D}_{\text{opt}}|} \sum_{c \in \mathcal{D}_{\text{opt}}} I\left[ \max(F(c_T)) < 0.05 \right].
\end{align}

\end{document}